# There Goes the Neighborhood:
# Relational Algebra for Spatial Data Search

Jim Gray
Microsoft Research

Alexander S. Szalay
Gyorgy Fekete
William O'Mullane
Maria A. Nieto Santisteban
Aniruddha R. Thakar
Johns Hopkins University

Gerd Heber
Cornell Theory Center

Arnold H. Rots
Harvard-Smithsonian Center for Astrophysics

**Table of Contents**

# There Goes the Neighborhood:
# Relational Algebra for Spatial Data Search

Jim Gray[1], Alexander S. Szalay[2], Gyorgy Fekete[2],
Gerd Heber[3], Wil O'Mullane[2], Arnold H. Rots[4], Maria A. Nieto Santisteban[2], Aniruddha R. Thakar[2]
(1) The Johns Hopkins University, (2) Microsoft, (3) Cornell Theory Center, (4) Harvard-Smithsonian Center for Astrophysics
Gray@Microsoft.com,{Szalay, gyuri , WOMullan Thakar}@pha.jhu.edu, heber@tc.cornell.edu, arots@cfa.harvard.edu

## Abstract

We explored ways of doing spatial search within a relational database: (1) hierarchical triangular mesh (a tessellation of the sphere), (2) a zoned bucketing system, and (3) representing areas as disjunctive-normal form constraints. Each of these approaches has merits. They all allow efficient point-in-region queries. A relational representation for regions allows Boolean operations among them and allows quick tests for point-in-region, regions-containing point, and region overlap. The speed of these algorithms is much improved by a zone and multi-scale zone-pyramid scheme. The approach has the virtue that the zone mechanism works well on B-Trees native to all SQL systems and integrates naturally with current query optimizers – rather than requiring a new spatial access method and concomitant query optimizer extensions. Over the last 5 years, we have used these techniques extensively in our work on SkyServer.sdss.org, SkyQuery.net, and TerraService.net.

## Categories and Subject Descriptors

H.2.1 Data Logical Design, H.2.2 Data Physical Design, H.2.8 Database Applications, J.2 Physical Sciences and Engineering, C.4 Systems Performance, E.1 Data Structures, E.1 Data Storage Representations,

## Keywords

Spatial search, databases, relational algebra

## 1. A Notation for Points and Regions

Spatial is special. Each spatial application seems to have some peculiar aspect that requires building a unique indexing method. In Astronomy, the special requirements are that most operations are done on the celestial sphere, and the typical search involves spatial, spectral, and temporal attributes of a high-dimensional space. The typical queries are *points-near-point*, *point-in-region* and *region-overlaps-region* – where regions are arbitrary polygons in space-time-spectrum coordinates.

The OpenGIS [OpenGIS] standard approximates these requirements, but it is not exactly right for them. It lacks the astronomical coordinate systems and projections (even though it has more than 70,000 geoid coordinate systems.) Astronomy has a legacy going back to the Phoenicians (minutes, seconds and degrees) that predate OpenGIS by a few millennia. Astronomers have accumulated quite a few other "standards" since then. So, we have borrowed concepts and terminology from OpenGIS, but followed the tradition of developing our own syntax and API.

Astronomy applications typically use a small subset of World Coordinate Systems that are defined for astronomical use. The most commonly used frames are either spherical or Cartesian coordinate systems; *equatorial* is aligned with the earth's rotation axis and equator, *ecliptic* is aligned with the earth's orbit, and Galactic is aligned with the Galactic plane. The equatorial coordinate system defined by the position of the earth's axis at the beginning of the year 2000 with longitude/latitude coordinates of *right ascension* and *declination* is most common. It is called the International Celestial Reference System (ICRS) or J2000. The community developed an XML schema for defining space-time regions [Rots]. In that standard, measured points have a location, error properties, and perhaps velocities. Regions of the sphere are described as the union of spherical polygons and their complements. Each polygon is in turn bounded by a set of arcs. Each arc is defined by its two endpoints (and the great circle passing through those endpoints) or by a small circle defined by the intersection of a plane with the sphere where the plane is described by its normal vector and its length. As with OpenGIS, regions may have a *buffer zone* that extends the region to include near-neighbors. Buffer zones are measured in arc-angles.

Humans never see the arcane XML syntax for regions; mostly they deal with graphical interfaces. But occasionally a compact linear syntax is wanted (analogous to the *well-known text representation* of OpenGIS [OpenGIS]. The rough BNF of this syntax is:

```
circleSpec  :=  CIRCLE J2000 ra dec radArcMin
             |  CIRCLE CARTESIAN x y z radArcMin
rectSpec    :=  RECT J2000 {ra dec}2
polySpec    :=  POLY J2000 {ra dec}3+
             |  POLY CARTESIAN { x y z }3+
hullSpec    :=  CHULL J2000 {ra dec}3+
             |  CHULL CARTESIAN { x y z }3+
convexSpec  :=  CONVEX { x y z d}+
regionSpec  :=  REGION { convexSpec }+
```

```
areaSpec    :=  circleSpec | rectSpec  | polySpec
            |   hullSpec | regionSpec
```

To give two examples, here is the definition of a 3 arc-minute circle centered at right ascension 30 degrees and declination 20 degrees.
```
 CIRCLE J2000 30 20 3
```
And the definition of a spherical triangle of the North-Eastern hemisphere is a sequence of ra, dec points
```
 POLY J2000  0 0  0 90  180 0
```

As with OpenGIS there is a natural Boolean algebra of these regions (union, intersection, negation.) There is also a natural spatial algebra for comparing points and regions. The typical queries are *point-proximity* ("What measurements are near this point?"), *point-polygon* queries ("What points are in this polygon?" and "What polygons contain this point?"), and *polygon-polygon* queries ("What polygons overlap or contain or are outside this polygon?"). The *simplification* query ("What is the "simple form" of this region definition?") also gives a test for empty regions. *Buffer-zone* queries ("What points or polygons are near this polygon?") are useful in many contexts. On the other hand we have not found a need for all nine Egenhofer OpenGIS spatial relationship functions (e.g. touches).

## 2. Working in 3D Avoids Spherical Geometry

Spherical metrics generally involve transcendental functions (*sine*, *cosine*, *tangent*,…) that are expensive to compute and that have singularities. It is computationally expensive to decide if a point is inside or outside a circle, or if two circles overlap. We use a *3D vector representation* to circumvent these problems. All points are represented as vectors on the unit sphere in Cartesian (J2000) coordinates. All circles are represented by the intersection of a plane with the unit sphere and a sign designating which side of the plane is inside the circle. The intersection of the unit sphere with the plane normal to vector $C = (x, y, z)$ of length $l$ *defines circle C*. Point $P$, represented as vector $(px, py, pz)$ is inside the circle if it is "above the plane," that is if $P \bullet C = x \cdot px + y \cdot py + z \cdot pz > l$. By going to 3-dimensions, point-in-polygon computations replace most transcendental computations with a few multiplies, adds, and a compare (Figure 1). We use this technique extensively.

Fuzz or boundary zones are important to many queries since all measurements are approximate and since one is often examining neighborhoods to look for clusters and local effects. The vector representation accommodates a fuzz of $\theta$ radians on the circle $C = (x, y, z)$ of length $l$ by replacing $l$ with $cos(acos(l)+\theta)$. The vector length is reduced by the cosine of the angle. To add a $\theta$ radian buffer to a polygon or region, just apply this transformation to each constraint of the region.

In what follows we describe three approaches to implementing these algorithms and the tradeoffs among the approaches. All the code is in the public domain and available at [SkyServer Regions].

## 3. The HTM approach

Virtually all spatial indexing techniques work on a hierarchical decomposition of space into bounding volumes that limit the search. Then a finer membership test is applied to all elements in the candidate boxes. The *Hierarchical Triangular Mesh* (*HTM*) first divides the sphere into 8 spherical triangles, and then builds a quad-tree recursively decomposing each triangle into 4 sub-triangles. Unlike many other spherical projection systems this one has the property that all triangles at the same level are within 42% of the area of all others and there are no singularities [HTM].

Each triangle can be named by a sequence $face, t_1, t_2, …, t_n$ where face is the index of the face of the major triangle, and each $t_i$ is an integer between 0 and 3 indicating which sub-triangle at that level has been chosen. This sequence is called the *htmID* of the triangle. Points can be described as tiny triangles – for example, a 20-deep mesh identifier on the surface of the earth corresponds to a triangle about 0.3 meters on a side, and a 30 deep mesh corresponds to a sub-millimeter-sized triangle on the geoid (0.3 milli arcseconds) and fits nicely in a 64-bit word. For most astronomy, a 20-deep htmID is adequate (0.3 arcsecond accuracy).

HtmIDs have a very useful property characteristic of space-filling curves: if *T1* and *T2* are HTM triangles, then *htmID(T1)* is a prefix of *htmID(T2)* iff *T1* contains *T2*. Storing the htmIDs in a Btree index will cluster nearby objects one another. All points or polygons within a triangle are located just after the parent triangle in the sorted list.

We built a library that, given a region as described in Section 1, returns a list of HTM triangles that cover that region [HTM]. We call this the HTM-cover. These triangles can be looked up in a B-tree and all points or polygons contained in those triangles are easily located. One can then run the "geometry filter" on those candidates to see if they qualify.

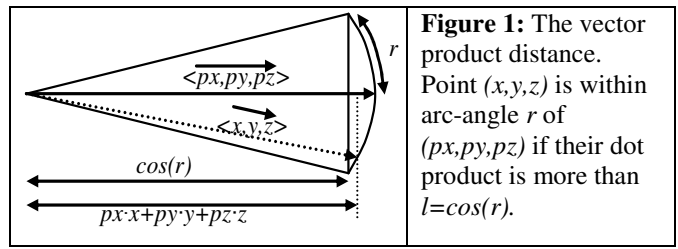

**Figure 1:** The vector product distance. Point $(x,y,z)$ is within arc-angle $r$ of $(px,py,pz)$ if their dot product is more than $l=cos(r)$.

The approximate logic of the HTM-cover routine is to consider each convex region in turn. For each region, construct a list of HTM triangles that intersect the region. Recursively divide each triangle on the region edge, trying to get a finer approximation to the region; discarding sub-triangles outside the region. The algorithm returns between 10 and 20 triangle ranges, which give an acceptable fraction of false positives (about 30%).

It is possible to spatially-enable almost any application by adding an HTM index and these HTM procedures to a pre-existing table. We call these extensions the *HTM-spine-schema*. We have added an HTM-spine-schema to a dozen astronomical databases by following these three steps:

First, htmIDs are represented as 64 bit quantities. The HTM code provides routines to convert between coordinate space and htmIDs with signatures something like this:
define function PointToHtmID (point varchar) returns htmID
define function HtmIIdToPoint(htmID bigint) returns varchar

Second, all point objects in the astronomy databases have their htmIDs computed when they are first ingested. An htmID field is added to each row and there is an HTM index on each such table T that allows very fast spatial searches.
create index T_htm on T(htmID, x, y, z)

Third, the htmCover table-valued function has the SQL signature:
define function htmCover (region varchar)
returns table (beginHtm htmID primary key, endHTM htmID)
The routine returns all points in the database that are included in the region.

That's all that is needed to spatially enable a table. The following query finds all points in table T within 3 arcminutes of the North Celestial Pole (there are 60 arcseconds in a degree and the pole is vector (1,0,0)):
select T.*
from T join htmCover('CIRCLE CARTESIAN 1 0 0 3')
  on T.htmID between beginHtm and endHtm
  and T.x*1 + T.y*0 + T.z*0 > acos(radians(3.0/60))

The last line of this query does the distance test using a dot-product. Figure 2 diagrams this "cosine" logic. This is an example of the careful geometry test following the coarse selectivity filter of the HTM mesh.

The above query is such a common operation that the spine schema implements a dozen table-valued fGetNearest and fGetNearby functions that return objects of a certain type within a certain radius of a given point. These functions use an HTM index to limit the search and then they filter the objects using the following equation to compute the actual distance between object *o* with celestial coordinates *o.x, o.y, o.z* and the point *x,y,z*:

$$DistanceInDegrees = degrees(2 \times asin(sqrt((o.x-x)^2+(o.y-y)^2+(o.z-z)^2))/2)) \quad (1)$$

This calculation in terms of *asin()* is more stable for very small distances (*acos()* is very close to 1 for small angles.)

The HTM design forms the basis for the SkyServer [SkyServer] and several other astronomical online databases. The performance can be roughly characterized as follows. On a 1 GHz Intel Pentium processor[2] the fixed cost of a null scalar function call in SQL Server is 31 μs, the cost of a null table-valued function call is 780 μs and the cost of a null external procedure call is 169μs. By comparison, the htmLookup computation takes 170 μs and the htmCover computation for a small circle or rectangle takes about 1.4 milliseconds. Much of this time goes into the linkage code between SQL and the HTM library written in C++. There is a substantial impedance mismatch between SQL and C++. SQL casts the HTM triangle-list into binary string and then into a SQL table.

Still, these routines are wrapped within SQLServer table valued functions that join the HTM triangles with the spatial data points and then run the geometry filter on each point. The base fGetNearbyObjXyz() runs in 6.7 ms for a 1 minute radius (28 objects returned, 35 objects examined. Other table-valued functions layered above fGetNearbyObjXyz() like fGetNearestObjXyz() or fGetNearbyObjEq(), add 3 ms to this cost (9.8 ms per call). So, most of the cost is in the procedure linkage, not in the HTM library.

With help from Beysim Sezgin and Peter Kukol of the Microsoft SQLServer group, we reimplemented the HTM libraries using the native virtual machine (the common language runtime) integrated with the next version of the product. This bypasses much of the SQL-HTM linkage cost. The resulting performance is described in Table 1. Clearly, the impedance mismatch is much reduced by integrating the virtual machine with the database. It also eliminates about 500 lines of very ugly glue code.

**Table 1:** Elapsed times (cpu milliseconds) of HTM functions using Transact-SQL or native virtual machine in SQLServer 2005.

|  | SQLServer 2000™ | | SQLServer 2005 + CLR | |
| --- | --- | --- | --- | --- |
|  | Null | Htm | Null | Htm |
| scalar | 0.03 | .17 | .05 | .09 |
| table valued | .5+.06R | 1.45 + .06R | 0.1 +.002R | 0.2 +0.003R |

In addition to the HTM implementation we have also implemented a HEALPix index [HEALPix] which gives a hierarchical iso-area and iso-latitude tessellation of the sphere and so are convenient for harmonic data analysis on the sphere (densities, integrals, spherical harmonics, Fourier transforms, etc.,). We are also flirting with an Igloo implementation [Igloo] which has similar properties and benefits.

---

[2] Unless otherwise noted, all measurements are done on a 1.1GHz Intel Pentium III processor.

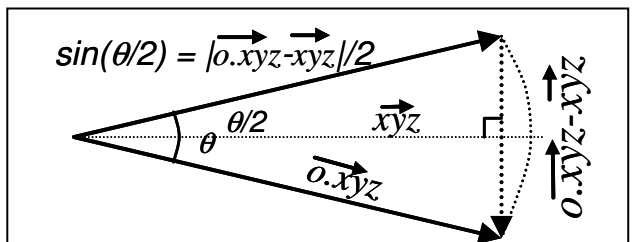

**Figure 2.** Equation (1) to computes arcangle distance between unit vectors *xvz* and *o.xvz*.

## 4. The Zone Approach

*Zones* are a way of bucketing two-dimensional spaces (or 2+D spaces) to give dynamically computed bounding boxes for queries.

### 4.1. The Problem – Going outside SQL is expensive

The basic problem is that SQL can evaluate equation (1) at the rate of about 170,000 records per second per cpu (5.6 µs per row) while the HTM functions run at about 170 records per second per cpu (6 ms per row to return ten nearby objects). This is a thousand-to-one performance difference. The previous section showed that the high cost of the HTM functions is a combination of the HTM procedures, the expensive linkage to SQL via external stored procedures (a string interface), and the use of table-valued functions. The HTM code uses about 1.5 ms and the other costs (linkage, string conversion, and table-valued function) are in the range of 4 ms. The linkage costs are much reduced with the integration of SQLServer 2005 with C#, but native execution will still have a substantial advantage.

### 4.2. The Basic Zone Idea

An alternative is to use SQL operations to limit the search rather than using the HTM procedures. Pushing the logic entirely into SQL allows the query optimizer to do a very efficient job at filtering the objects. In particular, the zone design gives a three-fold speedup for the table-valued functions. This same idea when applied to computing all the neighbors of each object in a 100 million object astronomy archive gives a 32-fold speedup, tuning a 14-day computation into a 9-hour job (see Section 4.3.)

The basic idea is to map the celestial sphere into *zones*, each zone is a declination stripe of the sphere with some *zoneHeight* (see Figure 3). The South Pole is in zone number zero. Then an object with a declination of *dec* degrees is in zone:

$$zoneNumber = floor((dec+90)/zoneHeight) \qquad (2)$$

There will be *ceiling(180/zoneHeight)* zones. The following code defines the zone table.

```
create table zone (zone int, objID bigint, ra float, dec float,
                   x float, y float, z float,
                   primary key (zone, ra, objID))
```

Notice that the primary key index on (zone,ra) clusters the elements of a zone's bounding boxes in Figure 3. The primary key index also makes (zone,ra) lookups very fast. The zone table is populated from table T *approximately* as follows.

```
insert into zone
    select floor((dec+90)/zoneHeight), ra, dec, x, y, z
    from T
```

### 4.3. Using Zones to Find Nearby Objects

If we search for all objects within a certain radius of point *(ra, dec)* then we need only look in certain zones, and only in certain parts of each zone. Indeed, to find all objects within radius *r* of point *ra*, *dec*[3], one need only consider zones between

$$maxZone = floor(\ (dec +90+R)/zoneHeight) \qquad (3)$$
$$minZone = floor(\ (dec +90-R)/zoneHeight)$$

and within these zones one only need consider objects o with

$$o.ra \quad between \quad ra-r \quad and \quad ra+r \qquad (4a)$$

(modulo *cos(dec)* and *ra wraparound* corrections in (4) below.)

This way of limiting the search is similar to the HTM approach but avoids calling an external procedure – it lets SQL do the math. The primary key on `zones` makes this lookup very fast, so that the resulting procedure has the performance given in Figure 4.

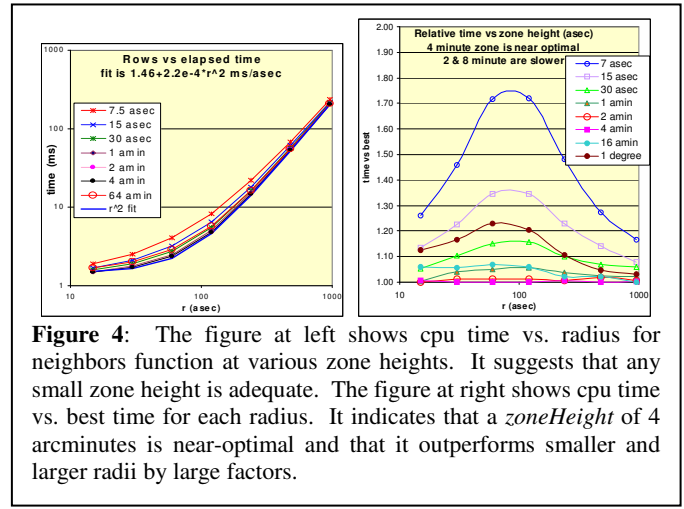

**Figure 4**: The figure at left shows cpu time vs. radius for neighbors function at various zone heights. It suggests that any small zone height is adequate. The figure at right shows cpu time vs. best time for each radius. It indicates that a *zoneHeight* of 4 arcminutes is near-optimal and that it outperforms smaller and larger radii by large factors.

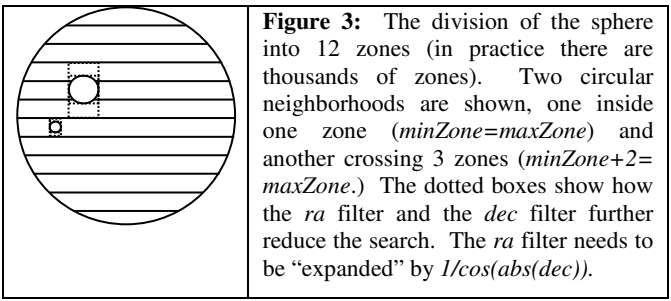

**Figure 3:** The division of the sphere into 12 zones (in practice there are thousands of zones). Two circular neighborhoods are shown, one inside one zone (*minZone=maxZone*) and another crossing 3 zones (*minZone+2= maxZone*.) The dotted boxes show how the *ra* filter and the *dec* filter further reduce the search. The *ra* filter needs to be "expanded" by *1/cos(abs(dec))*.

There are some nasty details that need a bit of extra mechanism. The biggest problem is that the sphere is round, so equation (4a) must be computed modulo 360°, and must be corrected for the fact that the right-ascension is "compressed" by *cos(dec)* as it moves away from the equator. So *ra* should be divided by *cos(dec)+ε* where ε is a tiny number added to prevent division by zero when *dec* is ±90°. Fortunately, equation (3) needs no correction, but equation (4a) should become

---

[3] We assume RA and DEC have been normalized to be in the ranges [0°, 360°] and [-90°, 90°] respectively.

*o.ra between     ra-R/(abs((cos(dec))+ε) and
                  ra+R/(abs(cos(dec))+ε)*     (4)

One other detail is that when one is near the prime meridian (*ra* = 0° or *ra* = 360°) then the "other end" of the range is nearby. This problem is solved by adding these neighbors into the zone as the *margins* with *ra* less than zero for the left margin and *ra* greater than 360° for the right margin. The main area is the ½ open interval [0°, 360°) and the margins must respect this ½ open property. Assuming a *MaxRadius* of 1° and *epsilon* is 1.0e-6:

```
insert into zone       -- right margin, notice +360 on ra⁴
 select  floor((dec+90) /@zoneHeight), ra+360, dec, x, y, z
 from T
 where ra >= 0
 and ra < @MaxRadius/(cos(radians(abs(dec)))+@epsilon)

insert into zone       -- left margin, notice -360 on ra
 select  floor((dec+90) /@zoneHeight), ra-360, dec, x, y, z
 from T
 where ra < 360
 and ra >= 360-@maxRadius/(cos(radians(abs(dec)))+@epsilon)
```

Now, equation (4) is actually correct and finds all neighbors within the zone. The full query to select the neighbors within @r of @ra and @dec from a zone is:
```
select objID
from  zone                          -- force the zone
where  zoneID = @zoneID             -- using zone number
 and ra between           -- quick filter on ra
    @ra - @r/(cos(radians(abs(@dec)))+ @epsilon)
   and @ra +@r/(cos(radians(abs(@dec)))+ @epsilon)
 and dec between @dec-@r    -- quick filter on dec
        and @dec+@r
 and 4*power(sin(radians(@r / 2)),2) > -- careful distance test
   power(x-@x,2)+power(y-@y,2)+power(z-@z,2)
```

This statement combined with the *minZone* and *maxZone* logic of equation (4) gives the performance described in Figure 4 for a table-valued function finding neighbors nearby a point. This is the 7x speedup over the HTM external procedures. The full statement handling this zone logic is:
```
select objID
from zone
where zoneID between floor((@dec+90-@r)/@zoneHeight)
            and floor((@dec+90+@r)/@zoneHeight)
and ra between @ra-@r/(cos(radians(abs(@dec)))+@epsilon)
       and @ra+@r/(cos(radians(abs(@dec)))+@epsilon)
and dec between @dec-@r
        and @dec+@r
and 4* power(sin(radians(@r / 2)),2) > -- careful distance filter
   power(x-@x,2)+power(y-@y,2)+power(z-@z,2)
```

---

⁴ In SQLServer, host language variables are preceded by a "@" character. So, here for example, @zoneHeight, and @epsilon are inputs or free variables in this SQL expression.

One can further accelerate the test by observing that `ra-@r` and `ra+@r` is too "fat" a band for any zone except the one holding the center of the circle. Figure 5 gives the equation for reducing the neighboring zone cell width.

### 4.4. Using Zones to Find Neighbors

Some queries want to compare several hundred million objects with all their neighbors. Astronomical searchers for gravitational lenses and for clusters are examples of such queries. To speed these queries the SkyServer precomputes the *Neighbors* table that lists all an object's neighbors within 30 arcseconds. This table averages about 9 neighbors per object; but, some objects have hundreds of neighbors and some have none. Using this materialized view is a thousand times quicker than searching for the neighbors each time -- 60 μs vs 6 ms per neighborhood.

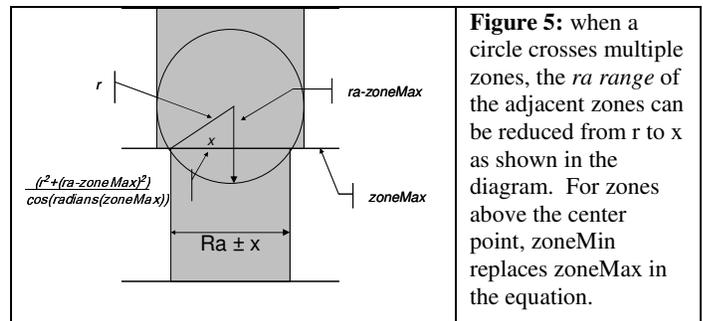

**Figure 5:** when a circle crosses multiple zones, the *ra range* of the adjacent zones can be reduced from r to x as shown in the diagram. For zones above the center point, zoneMin replaces zoneMax in the equation.

Computing the neighbors table using the fGetNearbyObjects function can take a long time: on the fifteen million object SDSS early data release, the computation took 56 hours – or about 74 neighborhoods per second. Fortunately, the computation was done only a few times during the load process and then used many times in queries. But, a speedup is needed as the SDSS database grows twenty-fold by 2007 and the naive computation grows to 2 months.

The computation is embarrassingly parallel and cpu-bound. Each object's neighbors can be computed independently. So a 30-node processor farm could do the 2 month job in 2 days. But, it makes sense to look for better algorithms.

The zone approach can bypass the stored procedure and get 30-fold speedup as follows. We can join each zone with itself and then with its north and south neighbor zones. These three joins all use the relational operators with automatic parallelism and with some very sophisticated optimizations. This bypasses much of the transact-SQL

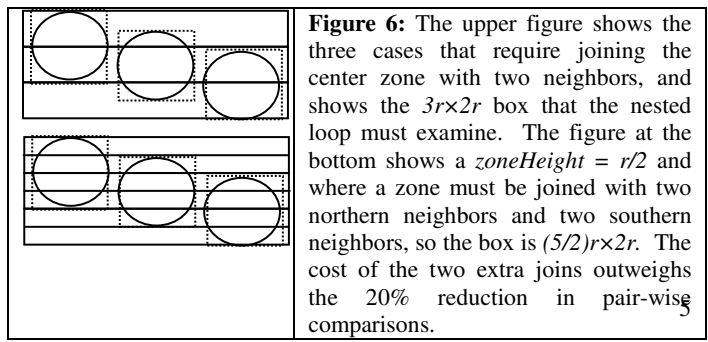

**Figure 6:** The upper figure shows the three cases that require joining the center zone with two neighbors, and shows the *3r×2r* box that the nested loop must examine. The figure at the bottom shows a *zoneHeight = r/2* and where a zone must be joined with two northern neighbors and two southern neighbors, so the box is *(5/2)r×2r*. The cost of the two extra joins outweighs the 20% reduction in pair-wise comparisons.

logic in the original algorithm.

The basic join to compute the neighbors is:

```
insert neighbors          -- insert one zone's neighbors
select o1.objID as objID,
    o2.objID as NeighborObjID,
         .. other fields elided
from zone o1 join zone o2    -- a nested loop join on
  on o1.zoneID-@deltaZone = o2.zoneID     -- zone,ra
  and o2.ra between
      o1.ra - @r/(cos(radians(abs(o1.dec)))+@epsilon)
   and o1.ra + @r/(cos(radians(abs(o1.dec)))+@epsilon)
where (  (o1.ra >= 0 and o1.ra < 360)  -- not both marginal
    or  (o2.ra >= 0 and o2.ra < 360))
  and o1.objID < o2.objID        -- do 1/2 the work
  and o2.dec between o1.dec-@r and o1.dec+@r
  and 4*power(sin(radians(@r / 2)),2) > -- careful distance
filter
power(o1.x-o2.x,2)+power(o1.y-o2.y,2)+power(o1.z- o2.z,2)
```

This is done for @deltaZone in {-1, 0, 1}. The insert-join above does only ½ the work, finding only objects where o1.objID<o2.objID[5]. To complete the neighbors table it is augmented with the mirror image of each pair (o2, o1). Then an (zone, ra) index is built on the resulting neighbors table.

This computation runs at 2.8k objects per second, computing the personal subset of the SDSS EDR in about a minute. The old algorithm took more than an hour on the same data and hardware. On the personal SkyServer (154k rows) the times for these steps on a warm database are:

Build zone table:   9.5 seconds
Join to zone  -1   10.5 seconds     generated 128,469 rows
Join to zone   0   16.5 seconds     generated 389,157 rows
Join to zone   1    9.4 seconds     generated 126,104 rows
Add mirror rows   10.7 seconds
Create index is     7.6 seconds
Total time         64.2 seconds     Total   1,287,460 rows

There is one surprise in the neighbors computation – it wants small zones (in particular *zoneHeight* = @r is optimal). Unlike the *nearby* computation (Section 3.3 and Figure 4) that works with a specific *ra, dec* in equation (4); the zone-join compares all objects in one zone to all objects in three other zones within the designated *ra* limits. That is, an object is compared with all objects in a box that is *2× radius* wide and *3×zoneHeight* high (the zone and its north and south neighbors as in Figure 6.) This means that "tall" zones result in quadratically more work Minimizing the *zoneHeight* minimizes work so *zoneHeight = radius* is optimal (½ arcminute is the radius for the neighbors table.) One might consider *zoneHeight* smaller than *radius*, but then one has to join with two or more northern and two or more southern neighbors as in Figure 6. These extra joins add extra costs that outweigh the savings in pair-wise comparisons.

---
[5] This optimization has two benefits: (1) It prevents marginal neighbors from being added twice (this might happen near the poles;) and (2) adding the mirror records, rather than computing them, speeds the computation by about 30% (see the following discussion of the cost of each phase).

### 4.5 Zone Summary

Using relational operators and a zoned-index to limit search speeds up the spatial proximity functions (GetNearbyObj et. al.) by about 3.4x and speeds the SDSS Neighbors table computation by about 32x. The neighbor table computation can be further accelerated by computing different zone pairs in parallel. Table 3 summarizes the speedups. One virtue of the zone approach is that is a way to implement spatial functionality in SQL without any proprietary extensions. It allows a completely portable library for points-near-point queries, and for some simple point in polygon and polygon overlaps queries.

|  | elapsed (ms/obj) | rate (obj/sec) | speedup |
|---|---|---|---|
| fGetNeighbors HTM | 14.5 | 69 | 1 |
| HTM Build Neighbors table | 13.5 | 74 | 1 |
| Zoned GetNeighbors | 1.7 | 578 | 8 |
| Zoned build Neighbors table | 0.2 | 2,406 | 35 |

**Table 3:** The times and speeds of computing the Neighbors with HTM or zone algorithms. The zoned algorithm is much faster. The neighbors-of-a-point speedup is 8:1 and the computation of all neighbors speedup is 35:1.

## 5. Representing Regions as Constraint Tuples

So far, the discussion has focused on point-in-polygon and nearby-points queries. Now we discuss ways to do algebra on regions and to do point and region queries with this representation.

### 5.1. Representing Regions

Section 2 explained that *spherical areas* can be represented as a set of positive and negative *convex-areas*. Non-convex areas may be composed as the union of several convex areas. Swiss-cheese areas with holes in them can be composed of positive and negative convex areas. Each convex area is defined by the intersection of the unit sphere with the interior of a 3D convex (possibly open) polyhedron, formed by these half-space constraints. The plane of a half-space constraint is in turn defined by a *normal unit vector v = (vx,vy,vz)* and *length l*. Point $P=(x,y,z)$ on the unit sphere is inside the circle if $(x,y,z) \bullet (vx,vy,vz) > l$. A point is inside a convex area if it is inside each of the half-space constraints. Figure 8 shows a complex convex area and also shows the dot-product test for "inside the half-space".

More generally, any *half-space H* of the N dimensional space *S* can be expressed as $H = \{x \varepsilon S \mid f(x) > 0\}$ for some function *f*. The intersection of a set of half-spaces $\{H_i\}$, defines a *convex* of points. $C = \{x \varepsilon S \mid x \varepsilon H_i \text{ for all } H_i\}$. A region *R* is the union of a set of convexes $R = \{x \varepsilon C_i\}$.

These ideas can be translated into relational database terms quite simply. A region is a name and an ID.

```
create table Region (
    regionID    int identity primary key,
    type        char(16),       -- short description
    comment     varchar(8000),  -- long description
    predicate   varchar(8000))  -- complied containment
    )       -- test see fRegionPredicate() below.
```

The region's convexes are stored as sets of 3D half-spaces together in a HalfSpace table:

```
Create table HalfSpace (
    regionID    int not null -- region name
        foreign key references Region(regionID),
    convexID    int not null,  -- grouping a set of ½ spaces
    halfSpaceID int identity(), -- a particular ½ space
    x   float not null,  -- the (x,y,z) parameters
    y   float not null,  -- defining the ½ space
    z   float not null,
    l   float not null,  -- the constraint constant
    primary key(regionID, convexID, halfSpaceID)
)
```

The following SQL query returns all the regions and convexes containing point @x, @y, @z.

```
select regionID, convexID from HalfSpace
    where @x *x + @y * y + @z * z <  l
    group by all regionID, convexID
    having count(*) = 0
```

This query groups all the half-spaces by their convexes. For each convex it asks how many of the half-spaces do NOT contain the point. If that answer is zero (count(*) = 0), then the point is inside all the convex's half-spaces and so is inside the convex and region.

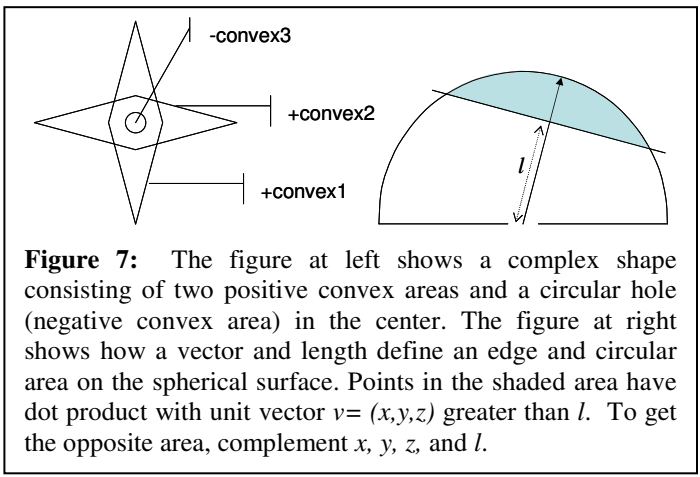

**Figure 7:** The figure at left shows a complex shape consisting of two positive convex areas and a circular hole (negative convex area) in the center. The figure at right shows how a vector and length define an edge and circular area on the spherical surface. Points in the shaded area have dot product with unit vector $v = (x,y,z)$ greater than $l$. To get the opposite area, complement $x, y, z,$ and $l$.

The key observation is that the HalfSpace table represents a region as a disjunct (*or*) of one or more convexes. Each convex is a conjunct (*and*) of its component half-spaces. The HalfSpace table is the disjunctive normal form representation of the region. The inverse (*not*) of a convex can be expressed through deMorgan's laws. If the convex $X$ has half-space constraints $a,b,c$, i.e. $X = a \bullet b \bullet c$, one can express its inverse $\sim X = \sim a + \sim b + \sim c$, where $\bullet$ represents *and*, $+$ is *or* and $\sim$ is *not*. It is advantageous to express this negation region as disjoint convexes. This can be achieved by expressing $\sim X$ as

$$\sim X = \sim a + a \bullet \sim b + a \bullet b \bullet \sim c$$

The negation of the half-space constraints is trivial, each component of the 4-vector $(x,y,z,l)$ is multiplied with $-1$.

The representation of spherical regions in terms of regions, convexes and half-spaces allows us to describe even the most complex "holes in holes" geometries. At the same time, the representation has some shortcomings: it ignores points exactly on the edge (the inequality is strict), but since our objects have finite extents with errors, this is not a problem.

### 5.2. Region Constructors

One can define a simple library for constructing regions and convexs. RegionNew creates a region:
regionID = **RegionNew**
            (type varchar(16), comment varchar(8000))
RegionNewConvex adds an empty convex hull to a region
convexID = **RegionNewConvex**(regionID int)
RegionNewConvexConstraint adds a half-space constraint to a region:
halfSpaceID = **RegionNewConvexConstraint**
                (regionID int, convexID int,
                 x float, y float, z float, l float)
To complete this we need the destructor:
**RegionDrop**(regionID).
Using these primitives, you can define complex regions. Indeed, regions can get *very* complex, so one needs a routine to discard empty convexes, merge neighboring convexes, and discard redundant constraints. RegionSimplify performs these operations.
        **RegionSimplify**(regionID)
RegionSimplify is quite complex and merits a separate paper. It considers each convex in turn seeing if it is null as follows: For every pair of planes, it computes the two intersection points of those two planes with the unit sphere. For $N$ planes, this is a list of $N^2$ points. A convex can form several patches on a sphere – each patch being a connected region—for example a cube can have eight triangular patches, one for each corner point. For each patch, we construct the list of arcs and points that meet in a vertex point at the boundary of the patch. We divide constraints not participating in any of the vertex points into two sets: those with $c<0$ (holes), and $c≥0$ (limits). First discard all limit constraints except the one with the largest $c$, i.e. the smallest circle. Discard all hole constraints where the anti-center point $(-x,-y,-z)$ is outside the patch. Simplify also merges adjacent convexes if they have the form $A = A.B+A.\sim B$. and discards a convex contained in another convex. The code is at [SkyServer Region].

These routines convert between the string representation and the half-space representation:
   string = **fRegionString**(regionID bigint)
and
   regionID = **RegionFromString**(string varchar(8000))
Other routines convert among coordinate systems.

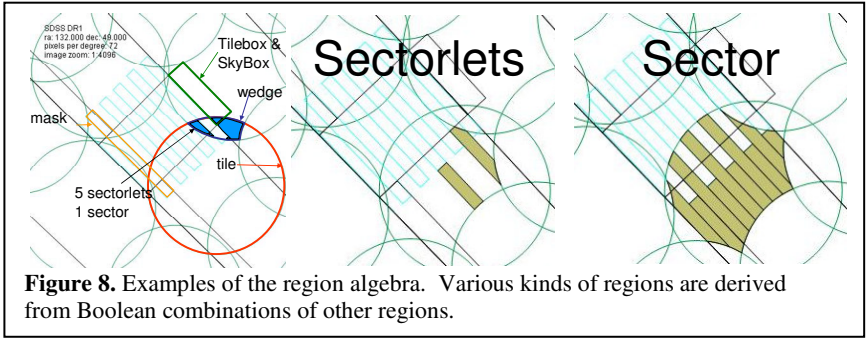

**Figure 8.** Examples of the region algebra. Various kinds of regions are derived from Boolean combinations of other regions.

### 5.3. Point-Region Queries

The regions representation is very convenient for point-in-region queries. The following routine returns all regions containing point (@x, @y, @z):
```
procedure RegionsOnPoint(@x float, @y float, @z float)
as select distinct regionID   -- return region name
   from HalfSpace              -- where
   where convexID in (         -- one of its convexes
        select convexID         -- contains the point, i.e.
        from HalfSpace          -- all half spaces contain the
        where @x * x + @y * y + @z * z < l
   group by all convexID       -- i.e., the point is not
   having count(*) = 0 )       -- outside any half space.
```

This query runs at the rate of 100K convexes per second per cpu (the inner loop is embarrassingly parallel). Conversely if one has many points and wants all the points in a certain region @regionID, the query is
```
create table Points (point ID int, x float, y float, z float).

define procedure PointsInRegion(@regionID int)
as select *                    -- return all points
   from Points p               -- in the area
   where exists ( select ConvexID  -- Where there is a
        from  HalfSpace h      -- convex in the region
        where  regionID = @regionID -- where no points
        and  (p.x*h.x + p.y*h.y + p.z*h.z) < h.l
        group by all ConvexID  -- outside a half-plane
        having count(*) = 0)   -- is zero (no outside points)
```

Again, this query runs at about 100K convexes per second. One can go about ten times faster by compiling the predicate as follows. First translate the predicate into an expression of the form:
   or (and ((p.x*a.x + p.y*a.y + p.z*a.z) >= a.l))
using
   predicate   = **RegionPredicate**(regionID)
Then combine the predicate with a select… where <predicate> and do an sp_execute of the resulting string. Something like the following:

```
declare @query varchar(8000)
set @query =
        'select * '              -- return all points in area
        'from Points p ' +       -- satisfying the predicate
        'where ' + fRegionPredicate(@regionID)  --
execute (@query)                 -- execute the query
```

This query runs at 1 µs per point or at the rate of a million half-space tests per second per cpu if it is not IO bound. Deriving the predicate costs less than 1ms and the fixed cost of executing the predicate on a small set of points is about 6ms (all this on a 1Ghz machine). So, if more than a thousand points are to be tested, the fixed cost of compilation is paid for by the speedup in point comparisons.

Unfortunately, SQL Server does not allow execute inside a function so one must use a temporary table and a stored procedure rather than using the more efficient table-valued variable.

### 5.4. Region Algebra

Once constructed, regions can be manipulated with Boolean operations.
regionID = **RegionOr** (regionID1 int, regionID2 int,
        type varchar(16), comment varchar(8000))
regionID = **RegionAnd** (regionID1 int, regionID2 int,
        type varchar(16), comment varchar(8000))
regionID =  **RegionNot** (regionID1 int,
        type varchar(16), comment varchar(8000))

These functions create new regions by adding a row to the Region table and many rows to the HalfSpace table. The 'OR' function just adds in all half-space rows from each of the two source regions with the new region name and with the convexID renumbered.

The "AND" predicate is more subtle. It intersects each convex from the first region with each convex from the second region. If there are N and M convexes in the original regions, then there will be NxM convexes in the conjunction. This is just an application of deMorgan's Law:
   *(A1 | A2) & (B1 | B2) = A1&B1 | A1&B2 | A2&B1 | A2&B2*

The negation predicate is by far the most complex. It needs to build a new set of convexes that draw a negative half-space from each of the original ½ spaces. The "and" and "or" were simple SQL statements. The negation required a recursive definition doing the Cartesian product of the (negation of) each half space in the first convex with the negation of all the other convexes in the region.

The algorithm uses RegionSimplify() to  simplify the resulting regions, discarding empty convexes.

The discussion here has been in terms of 3D space, because that is easiest to visualize. But, the algorithms and data structures apply to higher dimensions (more than 3D) by adding more parameters. They also apply to half spaces defined by higher-order polynomials (quadratic rather than linear equations). Further, the algorithms apply to non-Euclidian spaces like the surface of the sphere. The thing that makes the algorithms work is the triangle inequality inherent in any metric space.

### 5.4. Limiting Region Searches with Zones

When the number of regions gets large (more than 10,000 constraints) the region queries need to be restricted by some bounding box so that only a few thousand regions are considered in each query and so queries can be answered in milliseconds rather than seconds or minutes.

The standard approach is to use an R-Tree to represent bounding boxes [Samet]. Unfortunately, or perhaps fortunately, our database system does not support an R-Tree index. So, we have to do bounding boxes some other way. As it happens, the zoned approach fits with the SQL Query optimizer and so gives a spatial extension to SQL with no extra work by the vendor. This approach should work with any relational database system.

The zone idea (Figure 3) used earlier for point-point comparisons can be used for regions as well. Each region *R* has a bounding circle *radius*. If all radii were limited to the zone height (if all regions were smaller than twice the zone height), then we could just store a (zone, ra, dec, radius, regionID) tuple for each region in an index. Then when looking for regions overlapping a region R centered on @ra, @dec with @radius one would look in the neighborhood of zone number z = @dec/@zoneHeight. The zones to search would be limited to zones in the interval
    z ± ceiling(1+ @radius/@zoneHeight).
Within each of those zones, one would look at the box centered on @ra and look @buffer = @r + @zoneHeight to the left and right[6]:
    [@ra - @buffer , @ra + @buffer].

This is an efficient B-tree search. The zone search cuts the space by a factor of a thousand (if there are many thousand zones) and the @ra limit typically cuts the area by another factor of a hundred. So, careful geometry tests are only needed on few candidate regions.

The problem with this approach is that not all regions are of the same scale. Some are very large – covering ½ the sky, while most are tiny. The large regions have very large radius and so imply a huge buffer zone. To get useful bounding boxes we introduce the notion of a *zone-pyramid.* The index supports zones with granularities graded in powers of 2 having zone heights 1, 2, 4, 8,… until the coarsest zone covers the entire sphere. For the TerraServer, 200 meters is a reasonable *baseZoneHeight* (about 10 arcseconds), while for SkyServer 30 arcseconds seems to be the characteristic scale. These choices give about 100,000 base zones. There are about 16 scale levels to this system. Survey footprints that cover most of the sky are stored at the top-level scale while *masks* of bright stars saturated pixels are stored at the lowest scale.

```
create table zones (
        scale    int not null,  -- scale is 0,1,..., (2^n base
                                -- pixels per pixels at scale n)
        zone     int not null,  -- bucket that holds obj center
        ra       float not null, -- obj centerpoint (ra,dec)
        dec      float not null, -- == (lon, lat) for astro
        radius   float not null, -- obj bounding circle
        objId    bigint identity(1,1),  -- uniqueID
        primary key (scale, zone, ra, objID)
        )
```
Now, given an object with a certain bounding @radius centered on @ra, @dec, one can insert that object in the zones index by computing its scale (the zone height that is as big as this radius); and its zone (the band at that scale that contains @dec.)

```
set @BaseZoneHeight = 0.5/60.0  -- ½ arc second
set @scale = log2(ceiling(@radius/@BaseZoneHeight))
set @zone = floor(@dec/(@BaseZoneHeight* 2^@scale))
```

To search, we need to know what (scale, zone) pairs to examine and what buffer radius to use when looking in those zones. The following helper table-valued function returns between 15 and 40 zones depending on the radius – a larger radius returns more fine-grained zones[7].

```
fCandidateZones(@dec int, @radius int) returns
    @zones table (    scale         int,
                      Zone          int,
                      ScaleRadius   int,
             primary key(scale,zone))
```
Using that function, the following code finds objects within the circle centered at @ra, @dec with radius @radius:
```
select objID
from fCandidateZones(@dec,@radius) z join cells c
   on    z.scale = c.scale
         and z.zone = c.zone
         and  c.ra between @ra-z.ScaleRadius-@radius
                and @ra+z.ScaleRadius+@radius
where abs(c.ra-@ra)< c.radius + @radius
  and abs(c.dec-@dec) < c.radius + @radius
  and (c.ra-@ra)*(c.ra-@ra) + (c.dec-@dec)*(c.dec-@dec)
         < (@radius + c.radius) *(@radius + c.radius)
```

---

[6] Figure 4 shows an optimization of this idea.

[7] To work on the sphere (as opposed to the plane) the ra distance test must to be modified by cos(dec) and each zone has to have include duplicate objects at the margin (the sphere wraparound at the 0 meridian.) Both of these were explained in section 4.3 and are skipped over here. The implementation is at [SkyServer Region].

On a database of 1.3M randomly placed regions forming a 10-level pyramid, this query returns a list of overlapping regions in a few milliseconds while a brute force search takes several seconds. Tables 4 and 5 show the detailed performance of this design on a synthetic TerraServer tile database (the image pyramid) looking for areas that overlap a given area of radius 10 to 1000 pixels.

Based on the performance results of Table 5, it is quite reasonable to consider using this zoned-pyramid scheme to limit spatial searches for multi-scale regions. The region overlap performance is comparable to the performance of point queries (about 5ms) and is dominated by the cost of SQL's table-valued functions.

There is one caveat to this: if a region is long and narrow, then its bounding circle will be huge. Cosmic ray trails and drift scans have this property in the astronomy world and long interstate highways have this property in the geographic space. There are few drift scans and interstates so we can just ignore this problem for them (at the cost of many more false negatives), but cosmic ray, meteor, and satellite trails are very numerous. The solution in the SDSS is to segment these big objects into N smaller ones that all point to the same base object.

## 6. Summary

The Hierarchical Triangular Mesh code combined with SQL provides an HTM spine schema that makes it easy to add point and region queries to an existing database. With the advent of virtual machines integrated with database engines, the impedance mismatch between the spatial library and the database system is much reduced and the performance of the integrated HTM code improves by a factor of ten.

By using the zone idea (segmenting space into zone buckets and then segmenting zones by an offset), one gets performance comparable or better than the traditional (external procedures) HTM-SQL integration approach, largely because the zone approach allows the SQL optimizer to pick efficient plans and because we avoid the impedance mismatch – the zone code is all native to SQL. Adding a multi-granularity approach to the zone mechanism allows it to deal with areas of large dynamic range.

All this shows the simplicity and benefits of the HTM index. HTM uses a common representation for multi-granular searches that adapts to different scale factors with no extra mechanism. The code aims for an HTM cover of a small number of triangles which results in an area of the appropriate granularity.

Regions can be represented as unions of convex areas. This representation is very natural in a relational system and there are efficient ways for SQL to do points-in-region, regions-containing-point, and regions-overlap-region queries. This representation also allows a simple way to do Boolean algebra among regions. A routine from the HTM library is used to simplify regions and in addition regions can be simplified using geometric reasoning implanted as SQL set operations.

The region representation works well with both the zone and the HTM indexing mechanisms. The regions form a base layer, and these indices are different partitionings and indices of the underlying representation.

**Table 4:** The selectivity of the zone-pyramid scheme. Each row shows the result set cardinality vs radius for each successive filter clause in the SQL statement above. Zones cut the search space down to 10,000 items; the ra filter further reduces it by a factor of 100; and then the dec filter gives a further factor of 2. The fine ra filter does not contribute much. The "careful" geometry test has about π/4 positive answers (as would be expected of circles inscribed in bounding boxes.)

|  | Radius (in base units) | | | |
| --- | --- | --- | --- | --- |
|  | 10 | 100 | 350 | 1,000 |
| zone + scale filter | 7,968 | 7,968 | 10,004 | 19,589 |
| index ra filter | 70 | 75 | 109 | 316 |
| fine ra filter | 70 | 75 | 108 | 316 |
| dec filter | 41 | 46 | 84 | 275 |
| geometry filter | 31 | 35 | 62 | 218 |

**Table 5:** The performance of zone-pyramid search to find regions overlapping a region for various radii.

|  | Radius (in base units) | | | |
| --- | --- | --- | --- | --- |
|  | 10 | 100 | 350 | 1,000 |
| Count (objects) | 31 | 35 | 62 | 2216 |
| Time (milliseconds) | 5.0 | 5.3 | 5.4 | 7.8 |

We began by repeating the oft quoted phrase: "Spatial is special." The techniques here have proven very useful to us in building the SkyServer and other astronomical databases and have helped in the TerraServer. But, they each seem to form special cases rather than some grand unified approach.

The common themes are that it is possible to embed spatial concepts in a relational framework. When one does that, the SQL set-oriented language and optimizer is very convenient way to ask and answer the set-oriented queries typical of spatial applications. Surprisingly the SQL query optimizer, given either HTM or zone-pyramid indices, does a good job of efficiently answering these queries.

## 7. Acknowledgements